\documentclass[aapaper]{aa}
\usepackage{txfonts}
\usepackage[usenames,dvipsnames,svgnames,table]{xcolor}
\usepackage{graphicx}
\usepackage[draft]{hyperref}
\usepackage[hyphenbreaks]{breakurl}

\outer\def\gtae {$\buildrel {\lower3pt\hbox{$>$}} \over 
{\lower2pt\hbox{$\sim$}} $}
\outer\def\ltae {$\buildrel {\lower3pt\hbox{$<$}} \over 
{\lower2pt\hbox{$\sim$}} $}
\newcommand{\Msun} {M$_{\odot}$}

\newcommand{\Gaia}{{\sl Gaia}}

\usepackage{xspace}
\usepackage{amsfonts}
\usepackage{pifont}

\begin{document}
\title{Identifying Blue Large Amplitude Pulsators in the
  Galactic Plane using Gaia DR2: a case study}
\author{G. Ramsay\inst{1}}

\authorrunning{Ramsay et al.}
\titlerunning{BLAPs in Gaia DR2}
\institute{Armagh Observatory and Planetarium, College Hill, Armagh, BT61 9DG, UK\label{inst1}
\email{gavin.ramsay@armagh.ac.uk}}

%\date{Accepted: Oct 14 2018}

\abstract {Blue Large Amplitude Pulsators (BLAPs) are blue stars
  showing high amplitude ($>$0.2 mag) pulsations on a timescale
  of a few tens of mins.  They form a new class of variable star
  recently discovered using OGLE data. It has lead to a number of
  investigations searching for the origin of these pulsations. This
  short study presents the \Gaia\ DR2 data of ten BLAPs which have
  parallax measurements. We have dereddened their colours using \Gaia\
  DR2 data from the stars in their immediate field and find that six
  show absolute magnitude and intrinsic colour consistent with
  expectations, whilst four stars have a less certain
    classification. This work highlights the extra information which
  \Gaia\ DR2 data can provide to help classify those variable stars
  which do not currently have moderate resolution optical spectra and
  make searches for BLAPs in wide field high cadence surveys more
  systematic and robust.}

\keywords{Stars: oscillations -- distances -- white dwarfs -- ISM:
  dust, extinction}

\maketitle

\section{Introduction}

In the last few decades there has been a rapid increase in the number
of wide field photometric surveys which have led to the discovery of
hundreds of thousands of new Galactic variable stars as well as
transient events which have originated from nearby stars and
supernovae at cosmological distances.  Many of these discoveries have
been variable stars of well known types. For instance,
\citet{Alcock1998} used MACHO data to identify 1800 RR Lyr stars
located in the Galactic bulge which were used as tracers of Galactic
structure. More recently, the All-Sky Automated Survey for Supernovae
(ASAS-SN; \citet{Shappee2014}) has allowed the discovery of rare
objects, such as the ultra-compact AM CVn binaries
(e.g. \citet{Campbell2015}).

These surveys have also led to the discovery of new types of variable
star. One such discovery has come from the Optical Gravitational
Lensing Experiment (\citet{Udalski2015}, OGLE) and have been dubbed
Blue Large-Amplitude Pulsators (BLAPs). \citet{Pietrukowicz2017}
announced the discovery of more than a dozen blue stars which display
a periodic variation in the light curve on a period between 20--40 min
and with an amplitude of 0.2--0.4 mag. Their folded light curve is
very similar to the well known Cepheid and RR Lyr stars although with
a much shorter period.

\citet{Pietrukowicz2017} obtained moderate resolution spectroscopy of
one of their targets and determined a temperature $\sim$30000K. Lower
resolution spectra were obtained for a further three sources. They
have colours which are bluer compared to the upper main sequence. What
makes BLAPs interesting is that it is difficult for stars to pulsate
on such a short timescale {\it and} with a high
amplitude. \citet{Pietrukowicz2017} indicate that oscillations could
be driven in low-mass stars which have an inflated helium-enriched
envelope, while \citet{Romero2018} proposed they were the hot
counterparts of stars which are the progenitors of extremely low mass
white dwarfs. \citet{ByrneJeffery2018} tested whether post-common
envelope stars could be unstable and have the observed proporties of
the BLAPs and find that stars with 0.31\Msun could be their origin.

Given the interest of BLAPs to the field of evolved stars and that
wide field photometric surveys are likely to detect many more short
period variable stars, we have cross matched the available {\sl Gaia}
DR2 with the 14 BLAPs identified by \citet{Pietrukowicz2017}, with a
view to making further identification of BLAPs more robust.

\section{\Gaia\ DR2}

\begin{table*}
\caption{The first column shows the BLAP ID based on the number
  convention of \citet{Pietrukowicz2017}, where the optical period is
  also taken from, followed by the mean $G$ band mag; parallax and
  distance determined via a Bayesian approach and a scale parameter
  L=1.35kpc. There follows the $(BP-RP)$ (those in italics have been
  transformed from their $(V-I)$ colour), the absolute G magnitude
  uncorrected for reddening, $M_{G}$ and extinction, $E(BP-RP)$ (see
  text for details). The final columns, M$G_{o}$ and ($BP-RP)_{o}$ are
  the dereddened absolute G band magnitude and the dereddened
  $(BP-RP)$ colour.}
\label{sources}
\centering
\resizebox{\textwidth}{!}{
\begin{tabular}{llcccrccrr}
\hline
ID & Period & $G$ & Parallax         & d                   & $(BP-RP)$            & $M_{G}$                    & $E(BP-RP)$ & $MG_{o}$ & $(BP-RP)_{o}$ \\
     & (min)  & (mag)   &(mas)            &  (pc)               &                  &                       &        &          & \\
\hline
1  & 28.26 & 17.503$\pm$0.006 & 0.102$\pm$0.091  &  5330 (3730--10480) & 0.358$\pm$0.044  & 3.9$^{+0.8}_{-1.5}$  & 0.8   & 2.3$^{-1.5}_{+0.8}$ & -0.44$\pm$0.17 \\
3  & 28.46 & 18.915$\pm$0.010 & -0.117$\pm$0.450 &  3480 (1920--9210)   & 1.459$\pm$0.093  & 6.2$^{+1.3}_{-2.1}$  & 1.5   &  3.2$^{-2.1}_{+1.3}$ & -0.04$\pm$0.31\\
4  & 22.36 & 18.469$\pm$0.009 & 0.108$\pm$0.229  &  3770 (2330--9310)   & {\it 1.500$\pm$0.06}   & 5.6$^{+1.0}_{-2.0}$  & 1.4   &  2.8$^{-2.0}_{+1.0}$ & 0.10$\pm$0.29 \\
6  & 38.02 & 18.174$\pm$0.007 & 0.295$\pm$0.302  &  2930 (1760--8450)   & {\it 1.160$\pm$0.06}   & 5.8$^{+1.1}_{-2.3}$  & 1.0   &  3.8$^{-2.3}_{+1.1}$ & 0.16$\pm$0.21 \\
9  & 31.94 & 15.574$\pm$0.008 & 0.377$\pm$0.047 &  2650 (2250--3470)    & 0.626$\pm$0.028  & 3.5$^{+0.4}_{-0.6}$  & 1.4   & 0.7$^{-0.6}_{+0.4}$ & -0.77$\pm$0.28 \\
10 & 32.13 & 17.185$\pm$0.009 & 0.147$\pm$0.123  &  4370 (2970--9500)   & 0.389$\pm$0.055  & 4.0$^{+0.8}_{-1.7}$  & 1.4   &  1.2$^{-1.7}_{+0.8}$ & -1.01$\pm$0.29 \\
11 & 34.87 & 17.250$\pm$0.004 & -0.020$\pm$0.185 & 4670 (2960--10310)  & 0.231$\pm$0.028  & 3.9$^{+1.0}_{-1.7}$  & 1.3   & 1.3$^{-1.7}_{+1.0}$ & -1.07$\pm$0.26 \\
12 & 30.90 & 18.233$\pm$0.012 & -1.252$\pm$0.299  &  6170 (3860--12230) & 1.114$\pm$0.067  & 4.3$^{+1.02}_{-1.5}$  & 1.4   & 1.5$^{-1.5}_{+1.0}$ & -0.29$\pm$0.29 \\
13 & 39.33 & 18.931$\pm$0.008 & 0.719$\pm$0.409   &  1890 (1160--7430) & 1.505$\pm$0.061  & 7.56$^{+1.05}_{-3.0}$  & 0.9   & 5.8$^{-3.0}_{+1.1}$ & 0.60$\pm$0.19 \\
14 & 33.62 & 16.804$\pm$0.005 & 0.542$\pm$0.143  &  1920 (1460--4380)  & 0.014$\pm$0.059  & 5.4$^{+0.59}_{-1.8}$  & 0.7   &  4.0$^{-1.8}_{+0.6}$ &  -0.69$\pm$0.15 \\
\hline
\end{tabular}}
\end{table*}

The {\sl Gaia} Data Release 2 (DR2) on 25 April 2018 provided the
parallax of 1.3 billion stars down to $G\sim21$
\citep{GaiaBrown2018}. It also provided other parameters such as
proper motion and ($BP-RP)$ colour, which is derived from the prism
data. Of the 14 BLAPs identified by \citet{Pietrukowicz2017}, ten have
a parallax in \Gaia\ DR2. Every source was flagged as 'Variable' apart
from BLAP 4 which was recorded as having no variability information
available. Since the sources are not expected to be in the immediate
Solar neighbourhood, it is not surprising that the parallaxes (shown
in Table \ref{sources}) are small, with some even showing a negative
parallax (an indication of a source being at a high but uncertain
distance \citep{Astra2016a}).

We convert parallax from the {\sl Gaia} DR2 into distance following
the guidelines from \citet{BailerJones2015,Astra2016b} and
\citet{GaiaLuri2018}, which is based on a Bayesian approach. In
practise we use a routine in the {\tt STILTS} package
\citep{Taylor2006} and apply a scale length L=1.35 kpc, which is the
most appropriate for stellar populations in the Milky Way in general.
For most sources the lower limit for the distance is a few kpc with
upper limits nearing 10 kpc. In Table \ref{sources} we also note the
($BP-RP$) colour. For two sources, 4 and 6, no ($BP-RP$) colour is
given, so we take the $V-I$ colour reported in
\citet{Pietrukowicz2017} and convert to ($BP-RP)$ using the
transformations outlined in \citet{Evans2018}.

The BLAPs reported in \citet{Pietrukowicz2017} lie within a few
degrees of the Galactic plane: given they are all at distances greater
than a few kpc, the effects of extinction will be considerable and
need to be removed. There are a number of resources to determine the
extinction as a function of distance, including the 3D-dust maps
derived from Pan-STARRS1 data \citep{Green2018}, and the maps of
\citep{Capitanio2017} which were derived from a number of sources.
However, there are issues with these maps since they provide $E(B-V)$
rather than $E(BP-RP)$ and in the case of \citep{Capitanio2017} do not
extend further than 2kpc.

We therefore extracted sources from \Gaia\ DR2 within 0.25 degrees
from each source. Ideally we would then find the relationship between
distance and $E(BP-RP), A_{G}$ which are included in DR2 and derived
from the work of the {\sl Apsis} data processing pipeline (see
\citet{Andrae2018}).  \citet{Andrae2018} show that the relationship
between $E(B-V)$ and $E(BP-RP)$ is dependent on the intrinsic colour
of the source. We therefore extracted a subset from each dataset to
extract stars which had $(BP-RP)_{o}<$0.5 (which is what we expect for
the BLAPs) and determined the relationship between distance and
$E(BP-RP)$. In most cases, we found a reasonable agreement between the
values found using \Gaia\ DR2 data and the reddening maps mentioned
earlier after making an approximate conversion from $E(BP-RP)$ to
$E(B-V)$. As found by \citet{Andrae2018} we find the reddening $A_{G}
\sim 2 \times E(BP-RP)$. To estimate the total error on $(BP-RP)_{o}$
we assumed an error of 20 percent on $E(BP-RP)$ which was derived by
assessing the distance -- $E(BP-RP)$ maps. The dereddend values for MG
and $(BP-RP)$ are shown in Table \ref{sources}.

\begin{figure*}
  \hspace{2cm}
  \includegraphics[width=14cm]{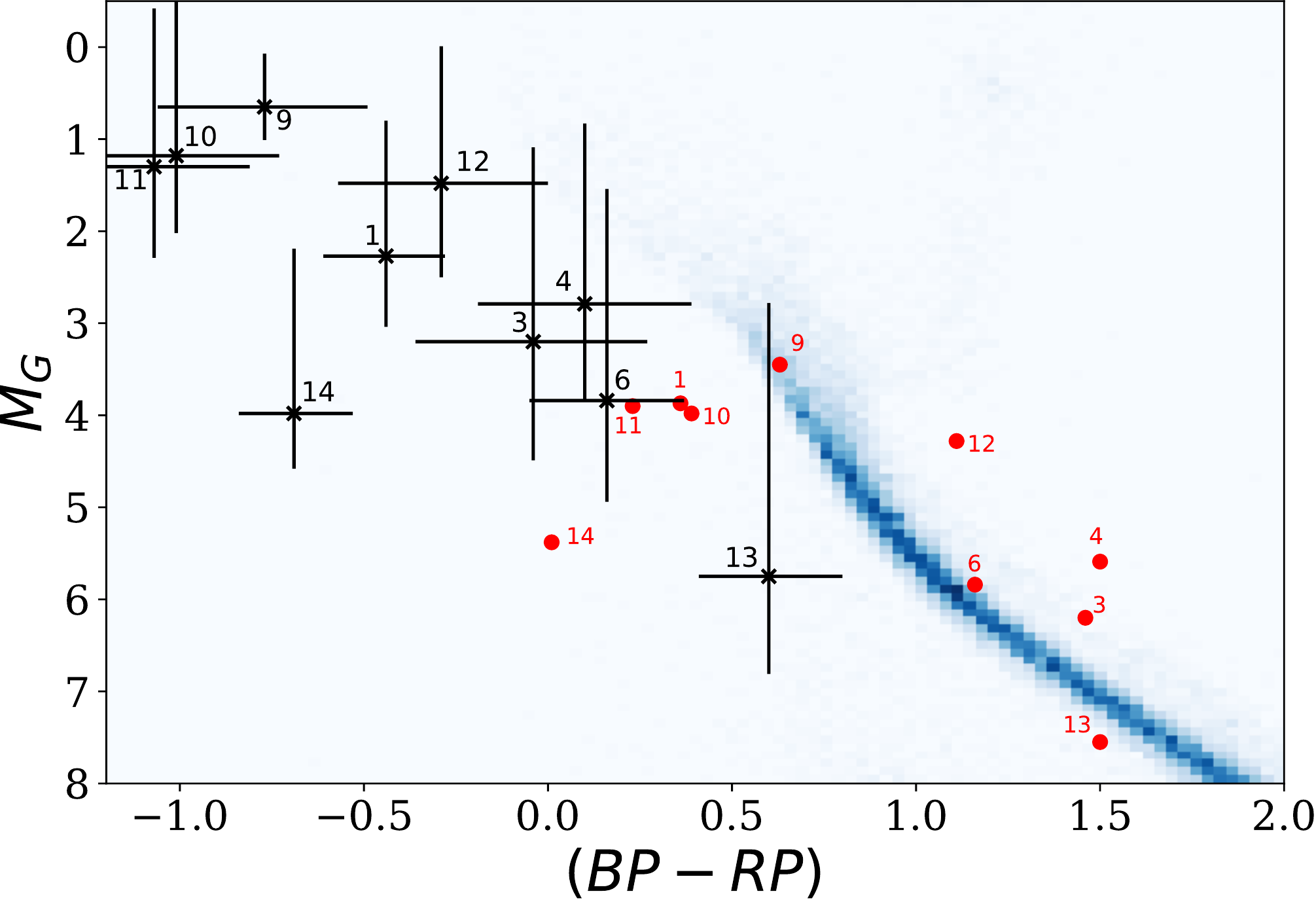}\hfill
  \caption{The unreddened ($BP-RP$), $M_{G}$ colour-absolute
    magnitudes for the BLAPs as identified by \citet{Pietrukowicz2017}
    are shown as red points, and the dereddened colour-absolute
    magnitudes are shown in black where the errors reflect the
    uncertainty on the distance, $(BP-RP)$ colour with an additional
    20 percent uncertainty on the extinction. The background shows the
    density of stars in the ($BP-RP$), $M_{G}$ plane of stars within
    150 pc which are assumed not to be significantly affected by
    reddening.}
\label{hrdered}
\end{figure*}

\section{Results}

The BLAPs identified by \citet{Pietrukowicz2017} are shown in the
((BP-RP), $M_{G}$) `HR' diagram in Figure \ref{hrdered} together with
a density plot of stars in all directions within 150 pc, which
we assume are little affected by interstellar reddening. Comparing the
observed colours (shown as red dots) with the dereddened colours
indicates that, as expected for stars in the Galactic plane, they are
significantly affected by reddening. However, although all the BLAPs
are bluer than the main sequence, there is a marked spread in the HR
diagram.

\citet{Pietrukowicz2017} presented moderate-resolution optical spectra
of star 1 and low-resolution spectra of stars 9, 11 and 14. They
determine a temperature for these stars within a relatively narrow
range ($T_{eff}\sim30000\pm$2500 K) and log $g\sim4.4\pm$0.2. These
stars together with stars 10 and 12 have an intrinsic colour
$(BP-RP)_{o}$<--0.5. Stars 3, 4 and 6 are redder ($(BP-RP)_{o}\sim$0.0)
and star 13 is significantly redder and fainter than those.

\citet{Andrae2018} present a detailed examination of how reliable
stellar temperature and reddening can be determined from \Gaia\ DR2
data. Although there is a degeneracy between temperature and reddening
over certain temperature ranges, they find that for known white dwarfs
with $T_{eff}\sim30000$ K, $(BP-RP)\sim$--0.4. Although there is a
considerable uncertainty in the distance (and to a lesser extent the
reddening) of the BLAPs studied here, we find that stars 1, 9, 10, 11,
12 and 14 have a implied temperature consistent with the temperature
of BLAPs derived from spectra and an absolute magnitude consistent
with theoretical predictions \citep{ByrneJeffery2018}.

In contrast, stars 3, 4 and 6 are redder and fainter. Comparing the HR
diagram of different classes of variable stars made using \Gaia\ DR2
data \citep{GaiaEyer2018} with Figure \ref{hrdered}, these stars lie
in the same part of the HR diagram as pulsating white dwarfs and
sub-dwarf B stars. Both of these variable star classes can show
periods on the same timescale as \citet{Pietrukowicz2017} found for
these stars (28.5, 22.4 and 38.0 min). On the other hand Star 13 is
fainter and redder and lies in the same part of the HR diagram as the
accreting Cataclysmic Variables (CVs).

\begin{figure}
  \hspace{0.2cm}
  \includegraphics[width=8cm]{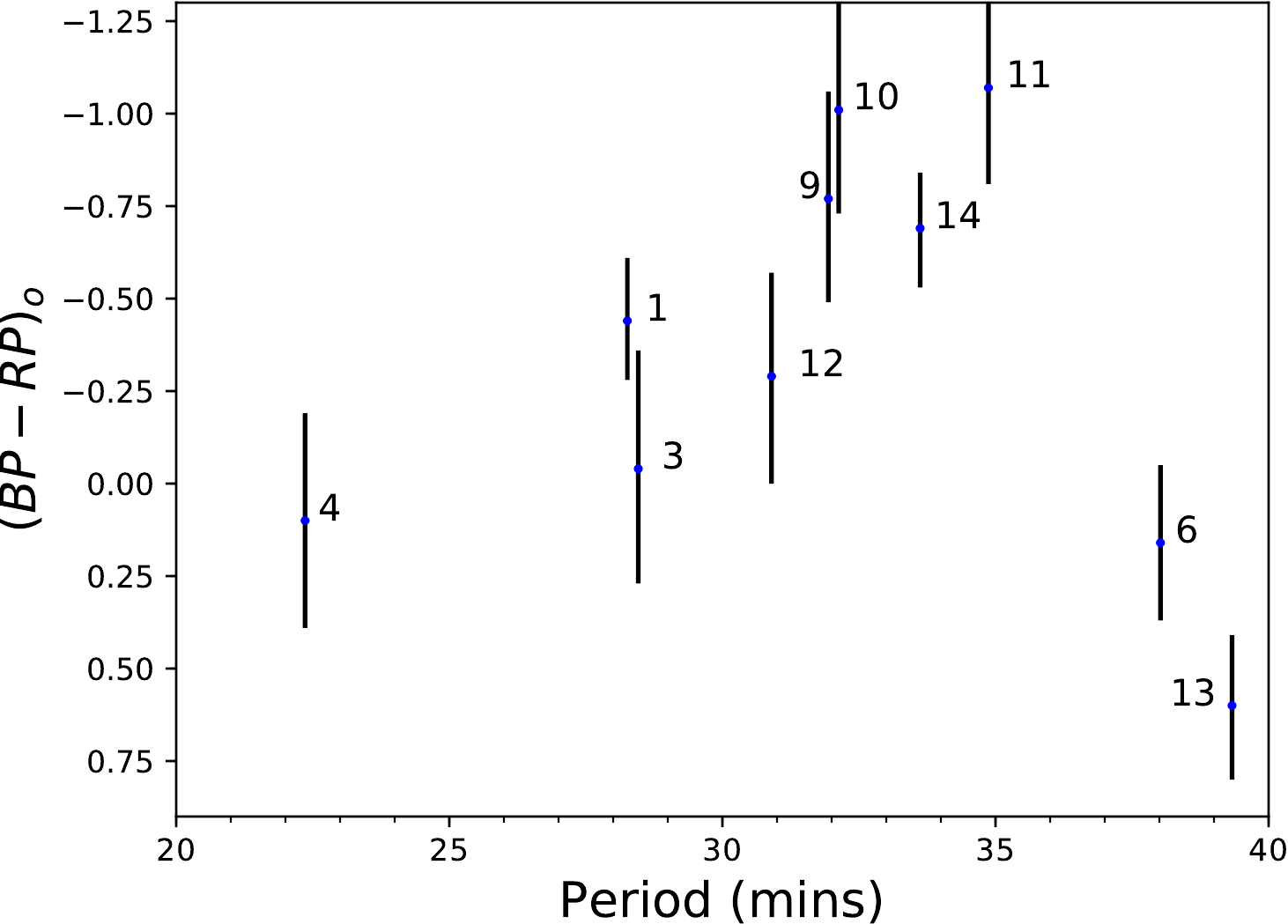}\hfill
  \caption{The dereddened $(BP-RP)_{o}$ colour for the BLAP sample as
    a function of period (taken from \citet{Pietrukowicz2017}).
    Figure \ref{hrdered} suggests that the physical characteristics of
    stars 4, 6 and 13, which are at the extreme ends of the period
    distribution, may differ from the other stars identified as
    BLAPs.}
\label{periodcolour}
\end{figure}

To investigate this further we compared the observed period of
modulation (see Table \ref{sources}) with $M_{G_{o}}$ and
$(BP-RP)_{o}$. Whilst there is no correlation between period and
$M_{G_{o}}$, we show in Figure \ref{periodcolour} the relationship between
period and $(BP-RP)_{o}$. Star 13 and 6 show both the longest periods
(39.3 and 38.0 min respectively) and the reddest colours. Star 4 has
the shortest period (22.4 min) with a colour similar to star 6. The
other stars, with the possible exception of star 3 which is slightly
redder, share a similar location in the distribution.  This adds
further evidence that stars 13, 4 and 6 have different properties
compared to the others.

\citet{ByrneJeffery2018} gives predictions for the dominant pulsation
period as a function of temperature for two masses and three types of
model.  For their prefered mass (0.31 \Msun) one of their models (the
`basic') predicts that shorter periods will be driven from hotter
stars.  The two other models (`standard' and `complete') have two
`loops' going through the region where pulsations are predicted to be
driven: one track where shorter periods have higher temperatures, and
the other with the reverse. Given the error on the $(BP-RP)$ colours
(Figure \ref{periodcolour}) we cannot favour either of the models.

Although there are many known types of blue variable stars, very few
show such a large amplitude on such a short period -- one of the
reasons why BLAPs have received such interest. For instance, although
sub-dwarf B stars can show pulsations on a similar timescale, they
show amplitudes typically of a few to tens of millimags. In contrast,
white dwarfs can show pulsation amplitude with a wide range. Indeed,
one of the first white dwarf pulsators to be discovered, HL Tau 76
\citep{Landolt1968}, was almost certainly discovered because of its
high amplitude (0.2 mag). CVs show variabilty on a range of period and
amplitude, including its orbital period, which can range from 5 mins
in the case of double degenerate binaries, to several hours for the
more typical. Star 13 shows a period of 39.3 min
\citep{Pietrukowicz2017} which would be too short for a hydrogen
accreting CV. However, even if the folded light curves shown in
\citet{Pietrukowicz2017} differ in detail, they all show a rapid rise
to flux maximum followed by a more gradual decline. They are very
similar to $\delta$ Sct, RR Lyr and $\delta$ Cep type variable stars
(albeit with much shorter period) but unlike the light curves of
hydrogen deficient CVs or indeed white dwarfs. They are therefore
likely to be stellar pulsators, although how similar to the BLAPs
remains to be seen.

\section{Conclusions}

In this study we have dereddended the colours of the BLAPs reported in
\citet{Pietrukowicz2017} using the reddening -- distance relationship
of stars in their immediate field. We find that six of the ten BLAPs
which have parallax information have an absolute magnitude and
intrinsic colours consistent with the temperature derived by those
BLAPs with optical spectra and theoretical predictions. Four stars
have properties which appear different and may be other types of
pulsating variables.  To determine their nature optical spectra of
these stars is a priority.

The large number of wide-field optical surveys searching for
transients and variable stars are identifying rare or new types of
variable star. For stars near the Galactic plane the effects of
reddening can make it difficult to determine the nature of a new
variable star, mainly because their intrinsic colours can be very
different to the observed colours.

The OmegaWhite survey is a high-cadence, wide-field photometric survey
which has covered 400 square degrees of the Galactic plane
\citep{Macfarlane2015} and is identifying thousands of new short
period variables \citep{Toma2016}. Whilst some of these fields have
VPHAS+ multi-colour data \citep{Drew2014} which helps to identify
their nature, for fields with high reddening the colours of compact
stars can be blended with the population of unreddened main sequence
stars.  \Gaia\ DR2 delivers information where the effects of reddening
can be removed at least to first order for each OmegaWhite field (one
square degree). It is therefore possible to select short period
variable stars with high amplitude which is more complete and robust
than would otherwise be the case.  This study highlights the
information which \Gaia\ DR2 can provide in characterising and
classifying variable stars which have no or only low resolution
optical spectra.

\section{Acknowledgements}

This work has made use of data from the European Space Agency (ESA)
mission {\it Gaia} (\url{https://www.cosmos.esa.int/gaia}), processed
by the {\it Gaia} Data Processing and Analysis Consortium (DPAC,
\url{https://www.cosmos.esa.int/web/gaia/dpac/consortium}). Funding
for the DPAC has been provided by national institutions, in particular
the institutions participating in the {\it Gaia} Multilateral
Agreement.  Armagh Observatory and Planetarium is core funded by the
Northern Ireland Executive through the Dept for Communities. I thank
Pasi Hakala for useful comments on an earlier draft of this paper and
an anonymous referee for useful comments on the submitted paper.

\end{document}